\documentclass[]{spie}  %>>> use for US letter paper
\pdfoutput=1
\usepackage{amsmath,amsfonts,amssymb}
\usepackage{graphicx}
\usepackage{multirow}
\usepackage[colorlinks=true, allcolors=blue]{hyperref}
\usepackage{rotating}
\DeclareMathOperator*{\argminA}{arg\,min}
\title{Information Content Approach to Trade Studies for SCALES}

\author[a]{Zackery Briesemeister}
\affil[a]{University of California, Santa Cruz, Santa Cruz CA 95060}
\author[b]{Steph Sallum}
\affil[b]{University of California, Irvine, Irvine CA 92697}
\author[a]{Andrew Skemer}
\author[c]{Natasha Batalha}
\affil[C]{NASA Ames Research Center, Mountain View, CA 94035}

\authorinfo{Further author information: (Send correspondence to Zackery Briesemeister)\\E-mail: zbriesem@ucsc.org}

\begin{document} 
\maketitle

\begin{abstract}

The advantage of having a high-fidelity instrument simulation tool developed in tandem with novel instrumentation is having the ability to investigate, in isolation and in combination, the wide parameter space set by the instrument design. SCALES, the third generation thermal-infrared diffraction limited imager and low/med-resolution integral field spectrograph being designed for Keck, is an instrument unique in design in order to optimize for its driving science case of direct detection and characterization of thermal emission from cold exoplanets. This warranted an end-to-end simulation tool that systematically produces realistic mock data from SCALES to probe the recovery of injected signals under changes in instrument design parameters. In this paper, we quantify optomechanical tolerance and detector electronic requirements set by the fiducial science cases using information content analysis, and test the consequences of updates to the design of the instrument on meeting these requirements.

\end{abstract}


\section{INTRODUCTION}
The Santa Cruz Array of Lenslets for Exoplanet Spectroscopy (SCALES) instrument will deliver 10-m class diffraction-limited thermal infrared low-/med-resolution spectra and imaging of low mass companions, cold brown dwarfs, circumstellar discs, and more. During the proposal stage, we identified the high-level requirements and overall specifications that lead all the design stages for instrument subsystems \cite{2020SPIE11447E..64S}. These fiducial science targets drive the optomechanical and electrical design decisions for each subsystem, defining requirements, modes of operation, relative position, orientation, quality, and general tolerances. Tolerances have a critical role in the manufacturing of instruments. Ultimately, the manufacturing accuracy of the instrument sets the limits of observability, and more importantly, characterizability of exoplanet atmospheres. This motivates our current study, in which we investigate how deviations in the optomechanics and electronics of the SCALES instrument will affect our ability to accurately characterize the atmospheres of exoplanets.

SCALES is equipped with low- and medium-resolution spectroscopic modes across six different wavelength ranges and nine different spectral resolving powers that can all be used for spectroscopy of exoplanets. In this paper, we seek to identify what the limits of these modes will be, in terms of exoplanet characterization, if the instrument were to deviate from the ideal design of SCALES.
The most rigorous way of accomplishing this is through atmospheric retrieval, which links atmospheric models to the data in a Bayesian framework\cite{2018haex.bookE.104M, 2020SSRv..216...82B}. While robust in its ability to derive non-Gaussian posterior distributions on atmospheric parameters of interest, it is computationally intractable given the diverse instrument phase space of SCALES and the possible atmospheric parameter space of known exoplanets. Instead, we treat this as an information content theory problem \cite{1949mtc..book.....S}, where one (or multiple) elements in an atmosphere and instrument model is perturbed. This allows determination of SCALES' ability to constrain physical atmospheric parameters (e.g. [$M/H$], $C/O$) as a function of various observational setups using "information loss", measured in bits, as a quantitative metric. We will model the effects of these perturbations with the SCALES simulation tool \cite{2020SPIE11447E..4ZB}. 

One might expect that if you perturb an optical element enough, the spectrograph simply will not work. So in between working and not working there is an extreme perturbation, such that no amount of exposure time would be sufficient to increase the information about model parameters with respect to priors. There is also a less extreme perturbation, such that you are limited by the amount of exposure time you can allocate to your target to gain sufficient information to satisfy fiducial science goals. This concept brings us to information content analysis, and restructures our investigation of tolerances in a way to answer the question: "Is the rate of information content per observation sufficient to satisfy the fiducial science case requirement for a given exposure time?"

\section{Information Content} 
The methodology we used for assessing information content of atmospheric observables from the modes of SCALES is based on optimal estimation theory \cite{2000imas.book.....R}. This theory has been used recently in two studies for optimizing JWST observations \cite{2017AJ....153..151B, 2017ApJ...835...96H}, and in a broad range of Earth and Solar System studies. We direct readers to (Rodgers 2000) for an in-depth description of information content analysis, as well as the notation used here. A brief summary of salient features of the information content analysis is included here for completeness. 

Information content $H$ is broadly defined as the change in entropy $S$ on making a measurement. In terms of probability distribution functions (PDFs) and Bayesian methods, it can be understood as the difference in entropy of PDF $P(\bf{x})$ and PDF $P(\bf{x}\vert \bf{y})$, where the state vector $\bf{x}$ describes the state of the atmosphere and the quantities which will be measured in order to retrieve $\bf{x}$ are represented by the measurement vector $\bf{y}$. In the case of SCALES, $\bf{y}$ is the spectrum measured. For our initial analysis, we simplify the state vector to be $\bf{x}$ = [$T(P)$, $[M/H]$, $C/O$, $g$], which neglects clouds. Here, $T(P)$ is the pressure-dependent temperature profile, $[M/H]$ is metallicity, $C/O$ is the carbon-to-oxygen ratio, and $g$ is the gravity. The information content $H = S[P(\mathbf{x})] - S[P(\mathbf{x}\vert \bf{y})]$ measured in bits describes the decrease in entropy of the probability that a certain state exists following a measurement.

The relationship between the measurement and the atmospheric state $\bf{y} = \bf{F}(\bf{x})$ is highly non-linear and depends on the instrument and spectral extraction methods used. However, a Taylor expansion of the model around some reference state is adequate for appropriately small deviations. This linearization sets the Jacobian matrix $K_{ij} = \partial F_i(\bf{x}) / \partial x_j$ as the appropriate operator, describing how sensitive the model is to slight perturbations in each state vector parameter. 

Under the assumption of Gaussian distributions, as done in (Rodgers 2000), 
\begin{equation}
    H = \frac{1}{2} \ln |\mathbf{\hat{S}}^{-1}\mathbf{S_a}|
\end{equation}
and 
\begin{equation}
    \mathbf{\hat{S}} = (\mathbf{K^TS_e^{-1}K + S_a^{-1}})^{-1}
\end{equation}
where $\mathbf{\hat{S}}$, $\mathbf{S_e}$ and $\mathbf{S_a}$ are the posterior error, data error and prior covariance of the state parameters, respectively. The quantification of information content relies on the construction of accurate Jacobian $\bf{K}$, and accurate estimation of prior and data error covariance. The parameter uncertainties are quantified in the diagonal elements of the posterior covariance matrix $\mathbf{\hat{S}}$.

\section{Modeling \& Retrieval Approach}
\subsection{Emission Spectra Models}

We use the PICASO code that enables computation of reflected light, thermal and transmission spectroscopy for exoplanets and Brown Dwarfs \cite{2019ApJ...878...70B}. For the analysis in this paper, we used self-consistent pressure-temperature profile from the Sonora 2018 model series for non-irradiated, substellar mass object \cite{2021arXiv210707434M}. Given the temperature-pressure profile of the atmosphere and the elemental abundances parameterized with metallicity, $[M/H]$, and $C/O$, the model first computes the thermochemical equilibrium molecular mixing ratios (and mean molecular weight) using the publicly available Chemical Equilibrium with Applications code (CEA\footnote{https://www.grc.nasa.gov/WWW/CEAWeb/}) \cite{mcb96}. The thermochemically derived opacity relevant mixing ratio profiles (H$_2$O, CH$_4$, CO, CO$_2$, NH$_3$, H$_2$S, C$_2$H$_2$, HCN, TiO, VO, Na, K, FeH, H$_2$, He), temperature profile, and planet bulk parameters are then fed into a emission spectrum model.

\subsection{SCALES Noise Models}
Sufficient reproduction of the processes involved in astrophysical detection is necessary to calculate $\mathbf{K}$ and $\mathbf{S_e}$. To be explicit in our modelling, we enumerate these processes here. 

In the detector model, the sensor can be treated as an array of pixels that take an input (a number of photons) and transforms these values to an output (a digital value in analog-to-digital units, or ADU). In performing this transformation, various sources of noise are added to the signal so that, given a camera's output in ADU, we can only make probabilistic inferences about the actual number of photons impinging on it. The simplest model would depend on five free parameters: the quantum efficiency $\eta(\lambda)$ [e${}^{-}/\gamma$], the read noise magnitude $\sigma_r$ [e${}^{-}_{rms}$], the dark current $\mu(T)$ [e${}^{-}$/s], the sensitivity (amplification of the voltage in the pixel from the photoelectrons) $K$ [ADU/ e${}^{-}$], and the bit-depth of the camera $k$. 

The noise generator of HxRG detectors from (Rauscher 2015) \cite{2015PASP..127.1144R} identifies and implements non-white sources of noise, which are added to our detector model. A model of the linearity of LMIRCam was used as the model for linearity in the simulation. The electric field from the AO simulations are converted it into units of photons from the irradiance of the field using the magnitude of the time-averaged Poynting vector of a linearly-polarized electromagnetic wave in free space. Photon shot noise from the source and background are added. ADU are treated as discrete integer units. Finally, the effects of readout direction, manifesting as crosstalk, are applied \cite{2018arXiv180800790G}.

The raw detector frames constructed by the simulation are then reduced into sky-subtracted data cubes $\mathbf{D}$ with the associated spectral covariance cubes $\mathbf{\Sigma}$ using the data reduction pipeline \cite{2018SPIE10702E..2QB}. A telluric calibrator star $\mathbf{M}$ is also simulated in a similar fashion. 

\subsection{Covariance propagation}

The proper approach to treating the PSF fitting as a weighted least squares problem becomes intractable due to the large covariance matrices that would need to be inverted. For context, integral field spectrograph data products are data cubes with ($n_{\lambda}, n_x, n_y$) dimensions. In principle, every point ($\lambda, x, y$) in this cube is correlated to every other point, so the appropriate covariance matrix would be ($n_{\lambda} n_x n_y \times n_{\lambda} n_x n_y$) matrix, or for SCALES low-resolution L-band, ($\sim 10^6 \times \sim 10^6$). Windowing in the spatial coordinates to select a region of interest would reduce this number significantly, but still greatly limits efficiency. 

Along with windowing to a 17 $\times$ 17 spaxel region around our target point spread function, we chose to make all spatial covariance components zeros so the matrix becomes sparse and a far more tractable problem \cite{2019AJ....157..244B}. We propose the data $\mathbf{D}$ at spatial coordinate ($j$, $k$) to be drawn from a multivariate normal distribution with mean $\mathbf{M} \mathbf{\beta}$ and covariance $\mathbf{\Sigma}$, for contrast $\mathbf{\beta}$ with respect to the calibrator $\mathbf{M}$. We then minimize the square Mahalanobis length to optimize the contrast spectrum.

\begin{equation}
\mathbf{\hat{\beta}} = \argminA_{\beta} (\mathbf{D} - \mathbf{M}\mathbf{\beta})^T\mathbf{\Sigma}^{-1}(\mathbf{D} - \mathbf{M}\mathbf{\beta}) = (\mathbf{M}^T\mathbf{\Sigma}^{-1}\mathbf{M})^{-1}\mathbf{M}^T\mathbf{\Sigma}^{-1}\mathbf{D}
\end{equation}
\begin{equation}
cov(\mathbf{\hat{\beta}}) = (\mathbf{M}^T\mathbf{\Sigma}^{-1}\mathbf{M})^{-1}
\end{equation}

These quantities are still contrasts with respect to the calibrator flux. Assuming the calibrator has the same exposure time and has not reached saturation, the contrast and covariance can be multiplied by the expected spectrum of the calibrator and its square, respectively, to represent the best-fit spectra $\mathbf{\hat{f}}$ and data error covariance matrix $\mathbf{S_e}$.

\subsection{The Jacobian}
The derivatives of the Jacobian $\mathbf{K} = \partial F_i/\partial x_j$ are calculated by centered finite-differencing scheme for each state vector parameter. We consider two different models for $\mathbf{F}$ in our approach. The first assumes $\mathbf{F}$ is only affected by the atmospheric physical model, the second assumes $\mathbf{F}$ is affected by both the atmospheric and the instrument model. The first scenario describes the case in which all instrument systematics can be removed during the data reduction process, and you are left with the raw atmospheric flux and the associated measurement errors. The latter considers the scenario that instrument systematics affect the ultimately flux measurements. Both models are depicted in Figure 1., where the pale blue Jacobians represent the first model and the $K$-, $L$- and $M$-band lines represent the second model. The instrument systematics do reduce the sensitivity at each band to elements of the state vector, and in a non-linear way. %Instead, the two spectra $\mathbf{f}[\mathbf{x} + \delta x_j]$ and $\mathbf{f}[\mathbf{x} - \delta x_j]$ with single-parameter perturbations $\delta x_j$ are propagated through the SCALES simulation to obtain $\mathbf{\hat{f}}[\mathbf{x} + \delta x_j]$ and $\mathbf{\hat{f}}[\mathbf{x} - \delta x_j]$.

%\begin{equation}
 %   K_{ij} \approx \frac{\hat{f}_i[\mathbf{x} + \delta x_j] - \hat{f}_i[\mathbf{x} - \delta x_j]}{2 \delta x_j}
%\end{equation}

\begin{figure}
    \centering
    \begin{minipage}{0.5\textwidth}
        \centering
        \includegraphics[width=0.9\textwidth]{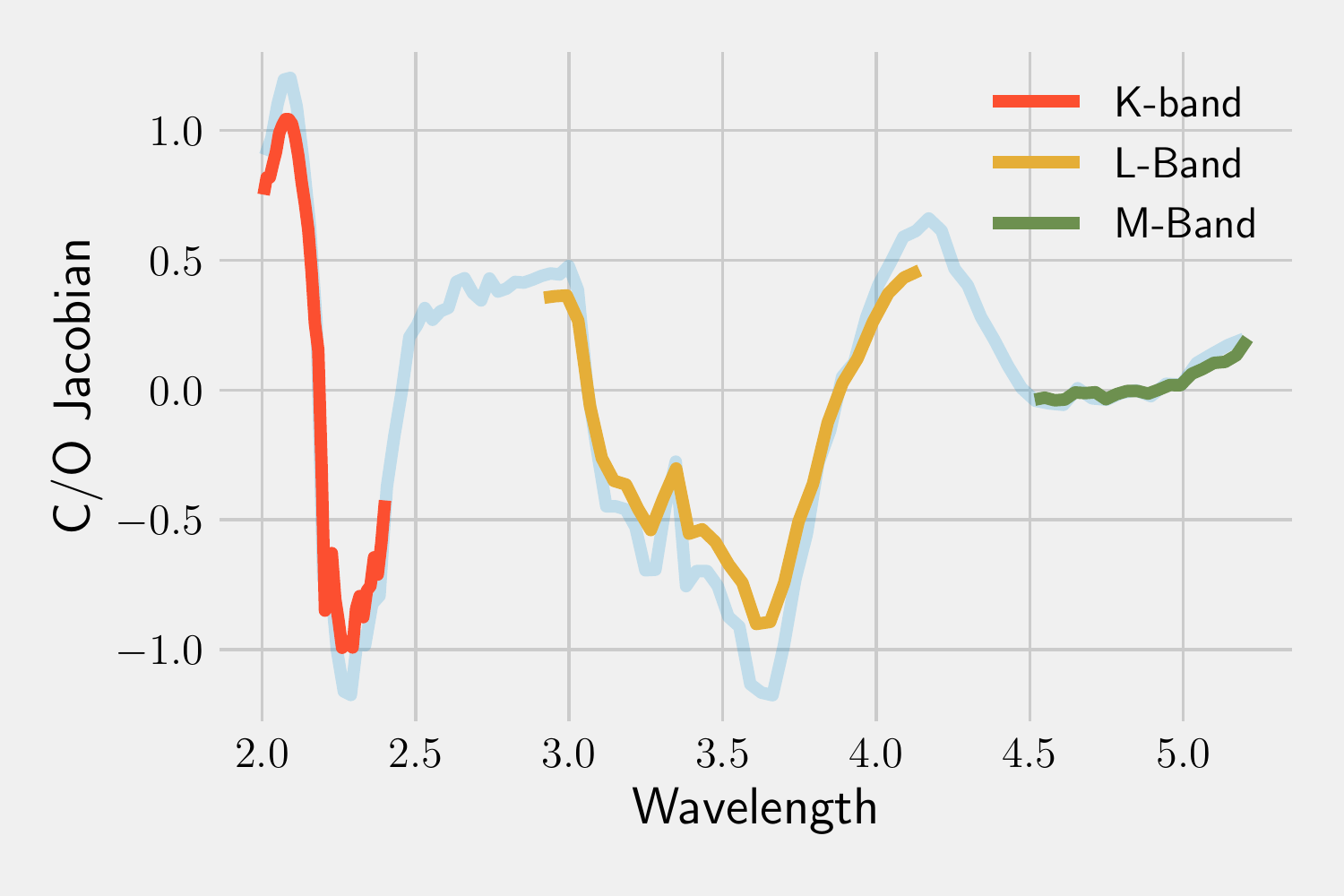} % first figure itself

    \end{minipage}\hfill
    \begin{minipage}{0.5\textwidth}
        \centering
        \includegraphics[width=0.9\textwidth]{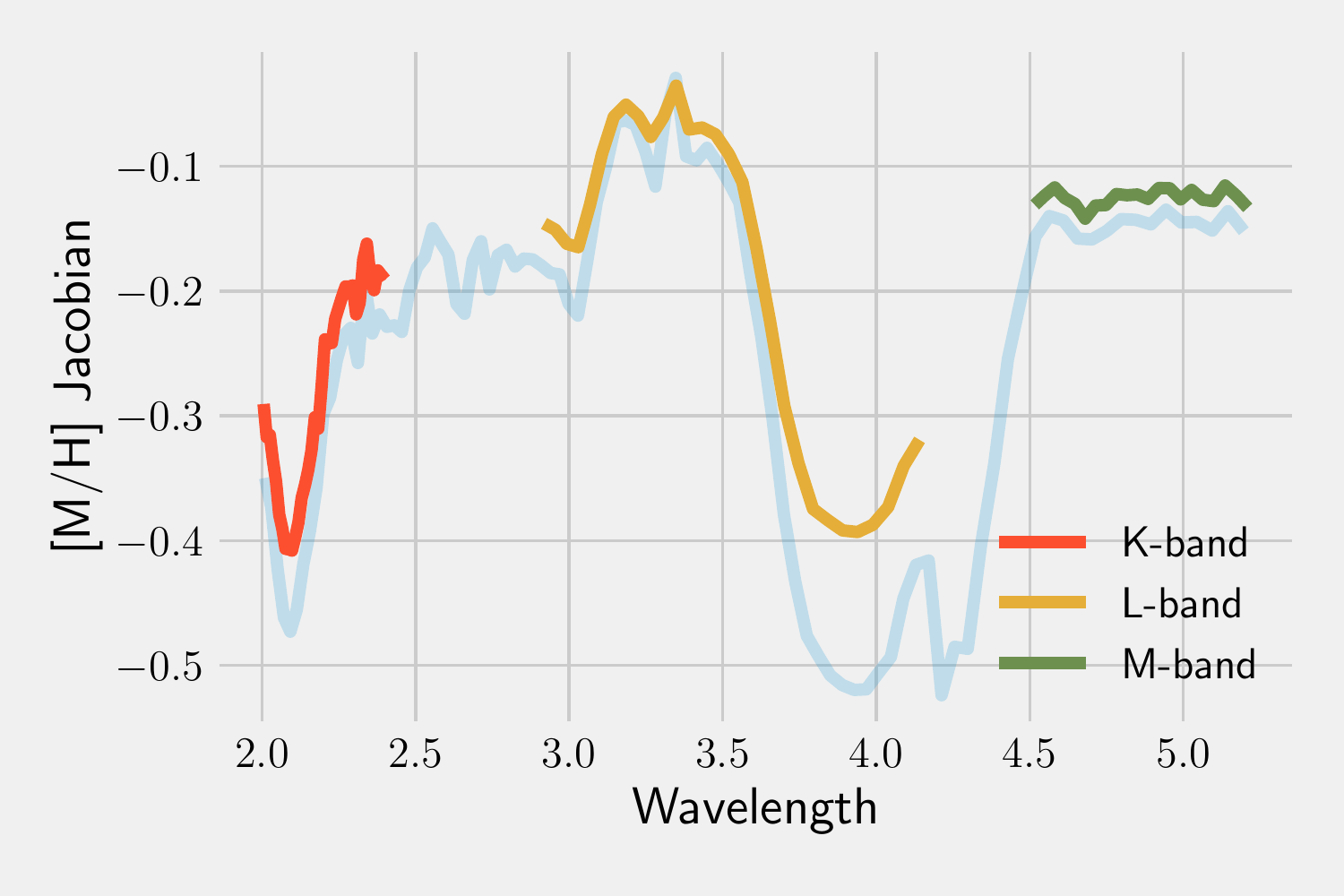} % second figure itself
        
    \end{minipage}
    \setlength{\abovecaptionskip}{8pt}
    \caption{The centered finite-difference scheme-derived Jacobians for $C/O$ and $[M/H]$ at $K$-, $L$-, and $M$-bands for a 1500K, $log(g)$ = 4, $[M/H]$ = 1, $C/O$ = .55. The deviations were $\pm$.001 in metallicity and $C/O$, and the Jacobians are scaled by $10^7$. The Jacobian in pale blue was calculated without passing the spectra through the SCALES simulator, binned at resolution of 100. The SCALES model is notably less sensitive at nearly all spectral bins to the hypothetical instrument capable of perfect recovery of the Jacobian.}
\end{figure}

\section{IC Analysis}
The following analysis explores information content for the two composition parameters, $[M/H]$ and $C/O$, that are used to parameterize elemental abundances that define the mixing ratios $\xi_i$. Other quantities [$T$, $g$] are of interest, but we wanted to be specific for our investigation. We set a naive prior $\mathbf{S_a}$ of $\pm 6$ $dex$ in both quantities, expressing essentially no prior knowledge of the atmosphere composition. 
 
SCALES has nine separate spectroscopic modes, each of which are sensitive to complementary and supplementary information content for various astrophysical signals. Naturally, one may expect there exist situations where observing in multiple modes would be advantageous from an information content context. The discussion in (Batalha \& Line 2017) presents arguments for why this may not necessarily be the case due to saturation of information. An in-depth exploration of using such multiple mode observations remains as work for future papers. In what follows we use information content to specifically focus on the 2-5.2 micron mode of SCALES, as it represents one of the more pathological observation modes due to the exceptionally bright sky background in $M$-band encroaching on bluer ends of neighbor spectra (Figure 2.)

\begin{figure}[]
\includegraphics[width=12cm]{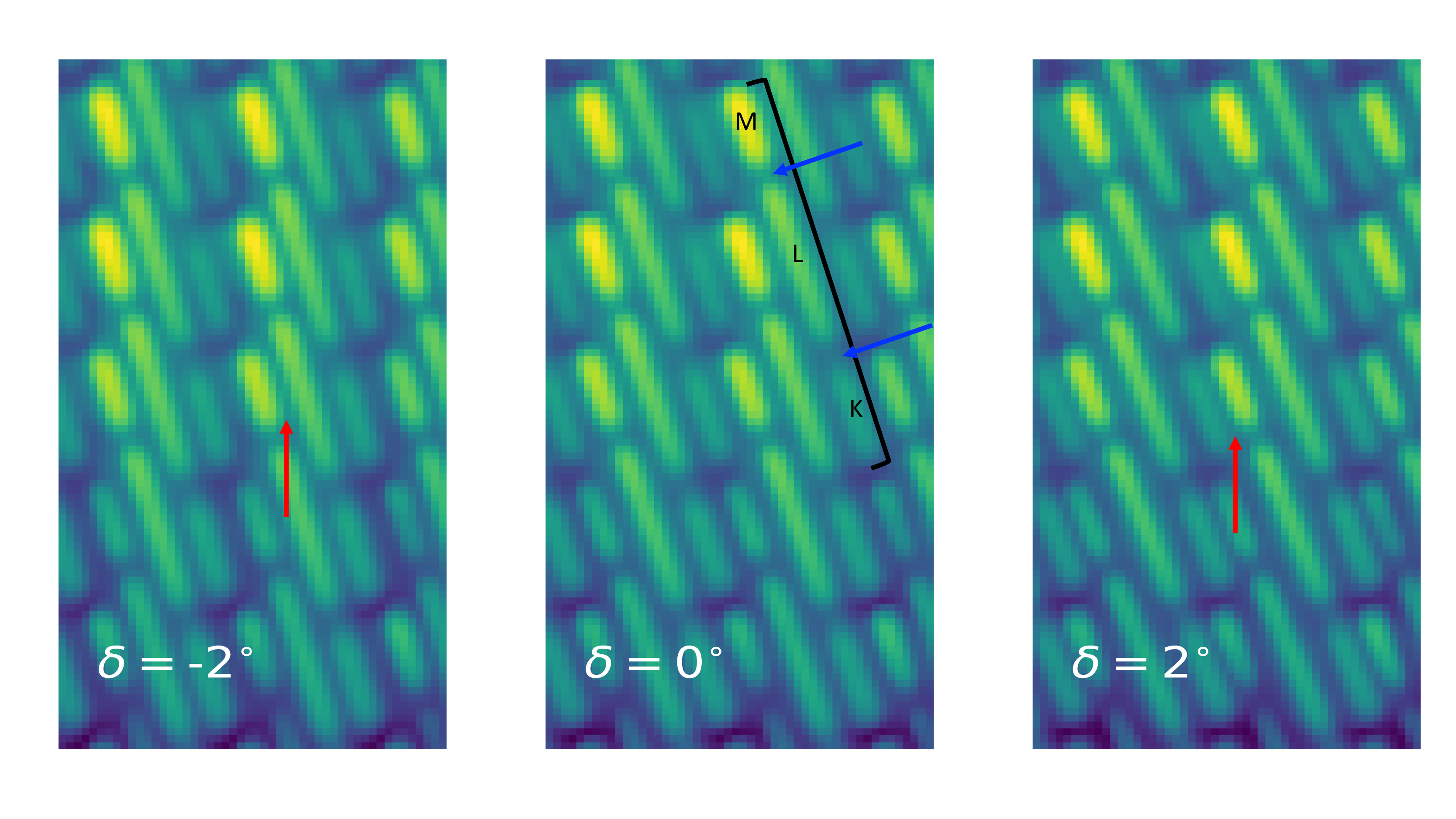}
\centering
\caption{The raw detector readout of a bright star in the 2-5 micron mode near the peak of the point spread function at three dispersion angles, set to tan${}^{-1}$(1/3) and  tan${}^{-1}$(1/3)$\pm 2^{\circ}$. The color scale is logarithmic. Detector systematics were removed for clarity. The anatomy of a single trace is distinguished in the middle panel, with a bright $M$-band lobe in the upper left, followed by a gap due to telluric water absorption marked in the blue arrow, a longer $L$-band lobe, another gap due to telluric water absorption marked in the blue arrow, followed by $K$-band light. The red arrows point out regions where the bright, variable thermal background in $M$-band lobes contribute to not only to flux, but also noise, in neighboring $K$- and $L$-band spectra, modifying the Jacobian $\mathbf{K}$.}
\end{figure}

\subsection{Case study: Misaligned Disperser}
This IC approach can be used to tolerance optomechanical design of SCALES. In this case study, we perturb the ideal model of SCALES, such that the angle of dispersion deviates from the fiducial value of tan${}^{-1}$(1/3). This manifests as a deflection from the ideal angle of the micropupils on the detector. This ideal angle was designed such that it maximizes the length and spectral resolution of the spectra as well as detector usage without the spectra overlapping. This case study will test how much the dispersion angle can be perturbed until the gain of information content is insufficient to satisfy the the fiducial science case within the exposure time.

\begin{figure}
    \centering
    \begin{minipage}{0.5\textwidth}
        \centering
        \includegraphics[width=0.9\textwidth]{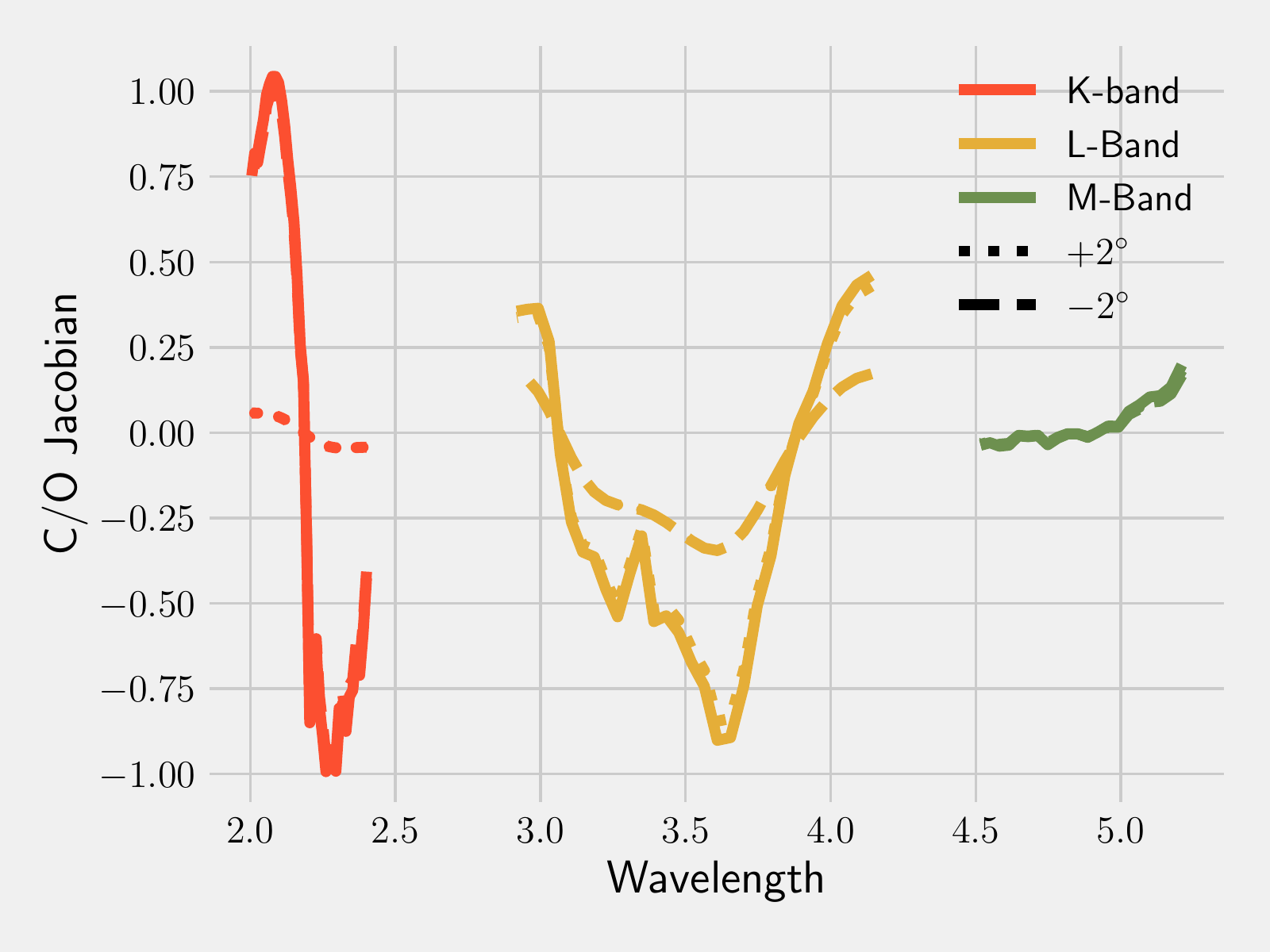} % first figure itself

    \end{minipage}\hfill
    \begin{minipage}{0.5\textwidth}
        \centering
        \includegraphics[width=0.9\textwidth]{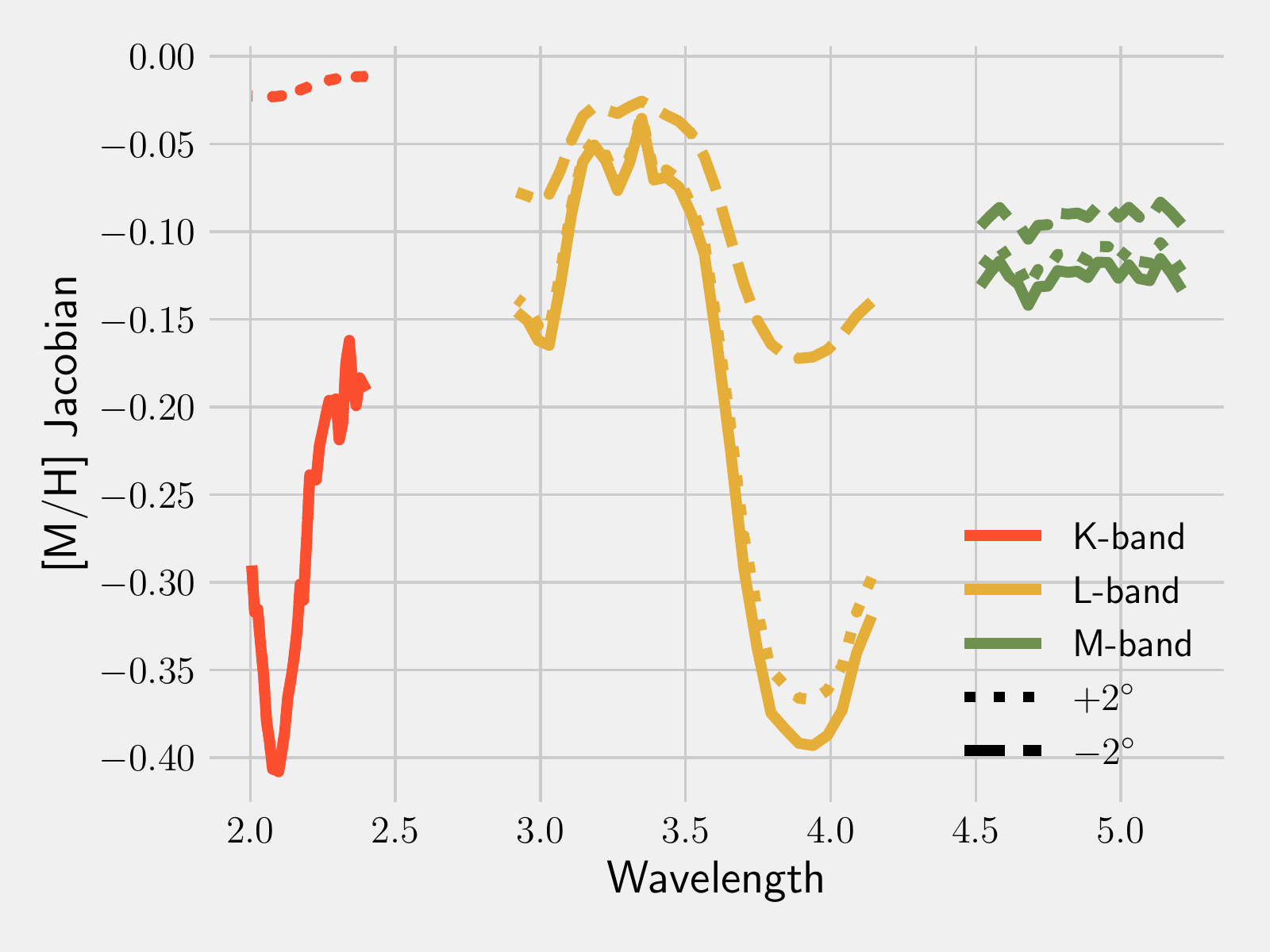} % second figure itself
        
    \end{minipage}
    \setlength{\abovecaptionskip}{8pt}
    \caption{The centered finite-difference scheme-derived Jacobians for $C/O$ and $[M/H]$ at $K$-, $L$-, and $M$-bands for a 1500K, $log(g)$ = 4, $[M/H]$ = 1, $C/O$ = .55. The deviations were $\pm$.001 in metallicity and $C/O$, and the Jacobians are scaled by $10^7$. These Jacobians are accompanied by the same measurements taken when the disperser is shifted by $+2^{\circ}$. The $K$- and $L$-band Jacobians are strongly diminished by impinging $M$-band light from neighboring spaxels.}
\end{figure}

In previous papers, we have used the simulation as an exposure time calculator to translate the fiducial science case requirements in terms of exposure time. In this case study, we set the requirement that we observe the signal-to-noise ratio of $C/O$ and $[M/H]$ to be greater than 5 in a 1500K brown dwarf at 100pc within two hours total integration time. The diagonal element of the posterior covariance matrix $\mathbf{\hat{S}}$ corresponding to $C/O$ and $[M/H]$ uncertainty is used to define the amount of information content necessary to meet this requirement. We then perturbed the model of SCALES by changing the angle of dispersion (Figure 2), and extracting the spectra assuming this new dispersion angle is known. Once the spectra begin to overlap, the overwhelming number of background photons in neighboring spaxels will begin to inflate the values of flux in the blue end of individual spaxels. This inflation rapidly deviates the extracted spectrum from the expected spectrum to the point where there is very little increase in information, relative to the prior (Figure 3).

According to our translation of this fiducial science requirement, and with all other degrees of freedom fixed, a realistic SCALES can deviate in dispersion angle from the ideal SCALES for all angles with information contents above 20.1 for $[M/H]$ and 20.4 bits for $C/O$ (Figure 4.). At angles more extreme than $\theta \in$ (18.2${}^{\circ}$, 18.7${}^{\circ}$), SCALES would not be capable of observing a source satisfying this requirement within the alloted exposure time.

\begin{figure}[]
\includegraphics[width=15cm]{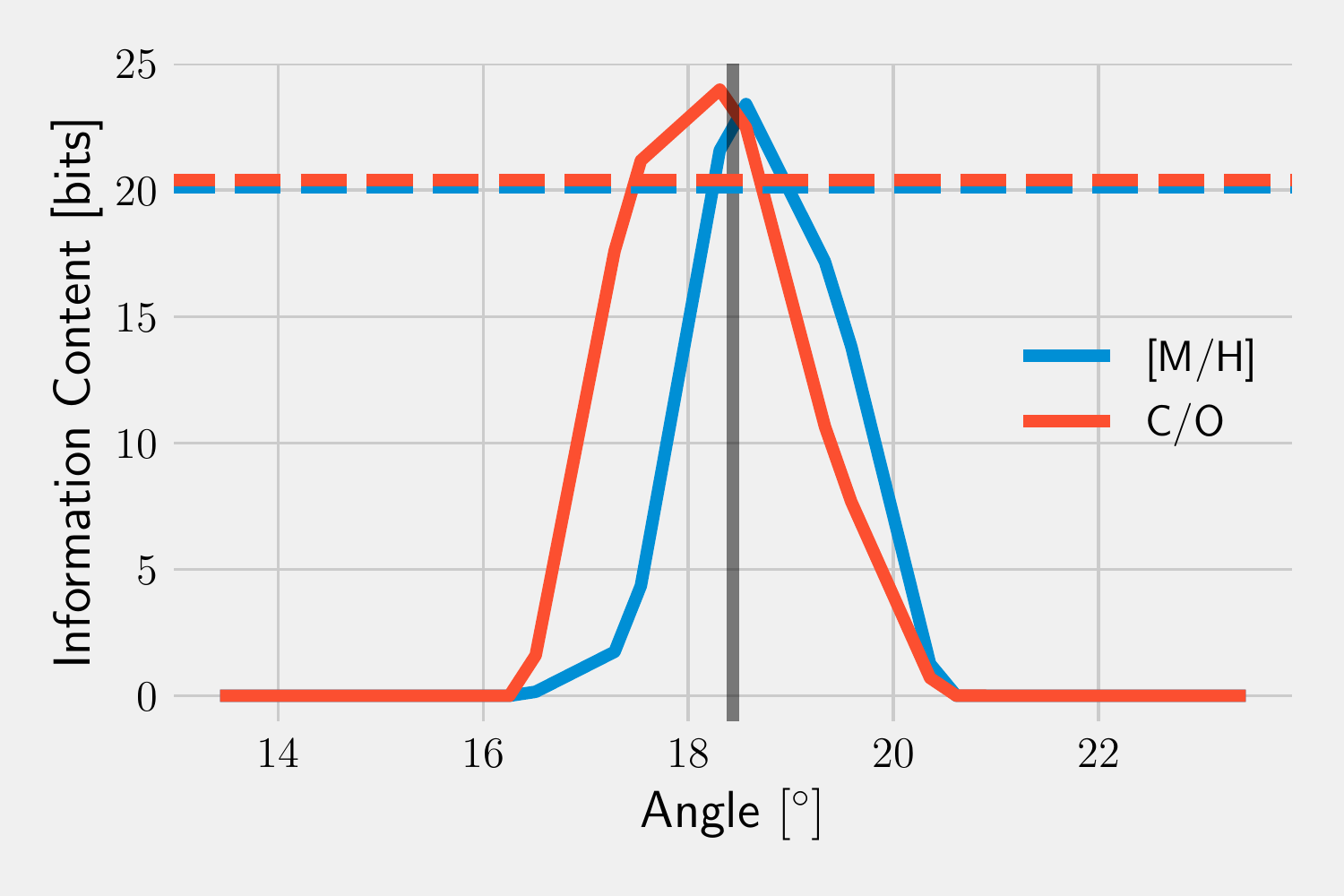}
\centering
\setlength{\abovecaptionskip}{8pt}
\caption{The information content in $C/O$ and $[M/H]$ for the fiducial science case described in the text at several dispersion angles. The vertical line denotes tan${}^{-1}$(1/3), and the respective horizontal lines corresponds to information content necessary to satisfy the fiducial science goal.}
\end{figure}

\section{Summary}

Using an emission spectra model, we computed how sensitive the SCALES instrument is to detecting changes in the state vector that define the models. These sensitivities make up the Jacobian, which allowed us to compute the information content of each spectrum. The information content, measured in bits, describes how the state of knowledge (relative to the prior) has increased by making a measurement. We applied this analysis to the case study of a particular optic that is perturbed in such a way to displace the intense thermal background at the red ends of these spectra onto neighboring, dimmer blue ends of spectra. The effect on information content was quantified in order to set a scientifically motivated tolerance of the angle of the dispersing element for SCALES. In doing so, we established a recipe to systematically follow to obtain tolerances other free parameters in the SCALES model. 

SCALES is expected to be deployed in 2025. After commissioning we will have a much better idea of what the inherent systematics are for each of the exoplanet spectroscopy modes. Until our knowledge of instruments improve, we can use these promising IC analyses to design optimized observing strategies for future proposals.

\acknowledgments 
We gratefully acknowledge the support of the Heising-Simons Foundation through grant \#2019-1697. ZWB is supported by the National Science Foundation Graduate Research Fellowship under Grant No. 1842400. This paper is based on work funded by NSF Grant 1608834.

% References
\bibliography{main} 
\bibliographystyle{spiebib}

\end{document}